\documentclass{emulateapj}
\usepackage{graphicx,amsmath,natbib,bm}

\usepackage[none]{hyphenat}

\usepackage{epsf}
\usepackage{epstopdf}

\input{epsf}
\usepackage{epsfig}

\usepackage{amsfonts}
\DeclareGraphicsExtensions{.jpg,.pdf,.png,.eps,.ps}

\def\msol   {\ifmmode{{\rm M}_{\odot} }\else{M$_{\odot}$}\fi}
\def\lsol   {\ifmmode{{\rm L}_{\odot}}\else{${\rm L}_{\odot}$}\fi}
\def\cii    {\ifmmode{{\rm C}{\rm \small II}}\else{C{\scriptsize II}}\fi}

\newcommand{\Ca}{\ensuremath{{\bf C}_\alpha}}
\newcommand{\CN}{\ensuremath{{\bf C}_N}}
\newcommand{\Cp}{\ensuremath{{\bf C}_{\rm p}}}

\newcommand{\T}{{\sf T}}
\newcommand{\vO}{{\bm O}}
\newcommand{\vp}{{\bm p}}
\newcommand{\va}{{\bm\alpha}}
\newcommand{\dOda}{\frac{\partial O}{\partial\alpha}}
\newcommand{\dOdp}{\frac{\partial O}{\partial p}}

\usepackage{color}

\usepackage[normalem]{ulem}

\slugcomment{}
\shorttitle{Dark Matter Subhalo Power Spectrum}
\shortauthors{Hezaveh et al.}
\begin{document}

\title{Measuring the power spectrum of dark matter substructure using strong gravitational lensing}
\author{Yashar Hezaveh\altaffilmark{1}$^,$\altaffilmark{2}$^,$\altaffilmark{3}, Neal Dalal\altaffilmark{4},
  Gilbert Holder\altaffilmark{3},  Theodore Kisner\altaffilmark{5}, \\ Michael Kuhlen\altaffilmark{6}$^,$\altaffilmark{7}, Laurence Perreault Levasseur\altaffilmark{8}}
\altaffiltext{1}{Kavli Institute for Particle Astrophysics and Cosmology, Stanford University, 452 Lomita Mall, Stanford, CA 94305-4085, USA}
\altaffiltext{2}{Department of Physics, Stanford University, 452 Lomita Mall, Stanford, CA 94305-4085, USA}
\altaffiltext{3}{Department of Physics,
McGill University, 3600 Rue University, 
Montreal, Quebec H3A 2T8, Canada}
\altaffiltext{4}{Astronomy Department, University of Illinois at
  Urbana-Champaign, 1002 W.\ Green Street, Urbana IL 61801, USA}
\altaffiltext{5}{Computational Cosmology Center, Lawrence Berkeley National Laboratory, Berkeley, CA 94720, USA}
\altaffiltext{6}{Theoretical Astrophysics Center, University of California, Berkeley, CA 94720, USA}
\altaffiltext{7}{LendUp, 237 Kearny St \#372, San Francisco, CA 94108, USA}
\altaffiltext{8}{DAMTP, University of Cambridge, Cambridge, CB3 0WA, United Kingdom}

\begin{abstract}  
In recent years, it has become possible to detect individual dark
matter subhalos near images of strongly lensed extended background galaxies. Typically, only the
most massive subhalos in the strong lensing region may be detected
this way. In this work, we show that strong lenses may also be used
to constrain the much more numerous population of lower mass subhalos
that are too small to be detected individually.  In particular, 
we show that the power spectrum of projected density fluctuations in
galaxy halos can be measured using strong gravitational lensing.  
We develop the mathematical framework of power spectrum
estimation, and test our method on mock observations.  We use our
results to determine the types of observations required to measure the
substructure power spectrum with high significance. 
We predict that deep observations ($\sim10$ hours on a single
  target) with current facilities can measure this power spectrum at
  the $3\sigma$ level, with no apparent degeneracy with unknown clumpiness in the
  background source structure or fluctuations from detector noise.
  Upcoming ALMA measurements of strong lenses are capable of placing strong
constraints on the abundance of dark matter subhalos and the underlying particle nature of dark matter.
\end{abstract}

\keywords{dark matter ---
gravitational lensing ---
galaxies: dwarf ---
galaxies: structure}

\section{introduction}

The abundance of substructure within the dark matter halos 
surrounding galaxies has been an area of intensive study for over a decade \citep[e.g][]{Moore:99, Klypin:99,dalal:02,Kravtsov:04,Simon:07,Strigari:07}.
Dark matter substructure in the present-day universe is
sensitive to the spectrum of primordial density fluctuations, generated
in the very early universe, implying that a precise quantification of
substructure can help constrain the physics of cosmic inflation \citep[e.g.,][]{Viel:04}.  In
addition, the microphysics of dark matter, such as its temperature or
the strength of its interactions, can also influence the structure of
dark matter halos and subhalos \cite[e.g.,][]{Lovell:12, Rocha:13, Lovell:14}.  Studies of halo substructure can
therefore probe multiple areas of fundamental physics.  
Comprehensive searches for faint and small satellite galaxies of the
Milky Way have revealed that the number of low-mass observable
satellites is significantly lower than what is predicted in CDM
simulations \citep{kravtsov:10}, referred to as the ``Missing Satellite
Problem''.  One possible explanation for this discrepancy is that,
perhaps, large numbers of dark matter subhalos exist but are not
observed because they are devoid of baryons, rendering them
effectively invisible.  If so, then the paucity of dwarf satellite
galaxies is a problem for galaxy formation models to address.
Another possibility is that the predicted subhalos simply do not
exist, pointing to new physics in the dark matter sector or inflation \cite[e.g.,][]{Lovell:12, Rocha:13, Lovell:14}.

Measuring the structure of dark matter on sub-galactic scales can
therefore shed light on the nature of dark matter and star formation
in dwarf halos.  An unambiguous characterization of the structure of
DM halos on sub-galactic scales requires a purely gravitational
detection method.  Gravitational lensing is an effective tool for
mapping out mass, even completely dark mass.  Strong
lenses, which produce multiple images of distant sources, are
sensitive to the presence of small-scale subhalos in lensing galaxies
\citep{mao:98}.  Subhalos can induce perturbations in nearby images
while leaving more distant (in angular separation) images unaffected.
An analysis of the differences between multiple images of a background
source can then reveal the presence of density perturbations near
images. \citet{dalal:02} used the anomalous flux ratios of multiple
images of 7 lensed radio quasars to constrain $f_{\rm sub}$, the fraction
of galaxy mass locked in subhalos, finding $0.6\%<f_{\rm sub}<7\%$ at
90\% confidence.  More recently, \citet{vegetti:10,vegetti:12}
showed that galaxy-galaxy strong lenses can be used to detect subhalos,
with two detections reported to date, resulting in $1.5\%<f_{\rm
  sub}<6.9\%$ at 68\% confidence for an assumed subhalo mass function
$dn/dM \propto M^{-1.8}$ \citep{xu:13}.  In addition,
\citet{Nierenberg:14} reported the detection of a subhalo in narrow-line emission of an optical quasar.
  These studies, however, have large
uncertainties due to small sample size (e.g., lensed radio quasars) and/or limited sensitivity to
subhalos (in case of extended source, e.g., galaxy-galaxy lenses).  Improving this measurement requires {\em both} a
significant increase in the sample of lenses, and a significant
increase in the sensitivity of each lens to the effects of
substructure. 
\citet{hezaveh:13a} suggested that the spatial and spectral resolution of ALMA  and the high signal to noise ratios of ALMA observations 
can allow us to detect of order one subhalo in bright lensed sub-mm lensed galaxies.
The analysis of the large number of newly discovered systems in this
sample \citep[e.g.][]{hezaveh:13b, vieira:13, bussmann:13} has the potential to yield a high
significance measurement of the abundance of subhalos with $M\gtrsim
10^8 M_{\odot}$. 

As discussed in \citet{hezaveh:13a}, only the few most massive subhalos can be
individually detected and characterized in typical strong lenses with extended sources. A
much larger number of subhalos are expected to exist at lower masses
(e.g. $\sim 10^6 M_{\odot}$), where various dark matter candidates
give rise to drastically different subhalo abundances.  Although
these subhalos cannot be individually detected, they can collectively
induce observable image perturbations.  These collective perturbations
allow the possibility of a {\em statistical} detection of the
population of low mass subhalos.  \citet{dalal:02} presented a method
for statistically constraining the properties of the DM subhalo
population using an ensemble of strong lensing systems.  Their method,
however, is computationally intensive, and therefore challenging to
apply to the large data sets that will be provided for extended lensed images by instruments such as ALMA.   

In this paper, we present an alternative method for constraining the
population of undetected subhalos using strong lensing data.  Instead
of modeling individual subhalos, we show that it is possible to
measure the power spectrum of the substructure density field by measuring 
the correlation of image residuals, after modeling the data with a lens with a smooth-potential. 
This technique allows us to probe the large population of unresolved subhalos with masses $\lesssim10^7 M_{\odot}$, by revealing their abundance and their average density profiles.
 In \S\ref{sec:power}, we discuss how the substructure power spectrum is related to the
properties of subhalos.  In \S\ref{sec:framework}, we present the
mathematical framework underlying our method.  Then in
\S\ref{sec:simulations} we apply our methods to mock observations, and
show how well the underlying substructure power spectrum may be
recovered.  Finally, we conclude in \S\ref{sec:discuss} with a
discussion of the implications of this work.
 
In all calculations we have assumed a cosmology with $\Omega_{\Lambda}=0.734$,  $\Omega_{m}=0.267$, and $h=0.71$, and assumed that the lens was placed at $z_d = 0.5$.

\section{The projected substructure power spectrum in dark matter halos}
\label{sec:power}

The distribution of substructure in ordinary galactic halos has been
studied extensively in the literature
\citep[e.g.][]{ViaLactea1,ViaLactea2,GHalo,Aquarius}.  N-body
simulations of $\Lambda$CDM cosmologies indicate that galactic dark
matter halos contain subhalos whose abundance approximately follows a
power-law distribution, $dN/dM \sim M^{-\alpha}$ with $\alpha \sim
1.9$, extending to the mass resolution limits of the simulations.
Internally, subhalos appear to have density profiles consistent with
tidally truncated NFW profiles.  The spatial distribution of subhalos
within their hosts may be somewhat less concentrated than the radial
distribution of dark matter, due to tidal stripping and destruction at
small radii  \citep{Diemand:04}.

The density fluctuations associated with substructure may be
considered as a random field, superimposed on the smoothly varying
background density profile of the host.  This field is not Gaussian.  However, because the subhalo mass function rises so quickly towards low mass, much of this non-Gaussianity is generated by the few most massive halos.  We expect to be able to detect these few massive subhalos individually, using direct lens modeling techniques (e.g. Hezaveh et al. \ 2013).  Below the detection sensitivity of these direct modeling techniques (e.g. $\lesssim5\times10^7 M_{\odot}$)
the number of subhalos is very large, reducing the non-Gaussianity of the density field. 

If we assume that the density field is Gaussian, we can fully
characterize it by its power spectrum.  A useful way to understand the
power spectrum is to use the halo model \citep{Cooray:12}.  This
decomposes the power spectrum into its contributions from subhalos of
different masses.  The ingredients of the halo model are (1) the
subhalo mass function, (2) the internal density profiles of the
subhalos, and (3) the distribution of subhalo locations.   Although
few subhalos are physically at small radii, they will occasionally
randomly project onto the strong lensing region ($r\sim 5-10$ kpc).
This implies that the projected number densities of subhalos will (on
average) be nearly constant as a function of projected $r$.  In other words, over the small 
regions probed by strong lensing,  we can assume that subhalos have a Poisson distribution with nearly
constant projected number density.  We can also neglect correlations
among subhalo locations.  In the language of the halo model, this
corresponds to neglecting the 2-(sub)halo term.  The reason it is safe
to neglect subhalo-subhalo correlations is that subhalos reside within
the tidal gravitational field of their host.  Subhalos that are not
gravitationally bound to each other will follow orbits whose relative orbital
phases wrap by order-unity angles within a few dynamical times, i.e.\
a timescale that is short compared to the Hubble time \citep{Chamberlain:14}.  Therefore,
although subhalos have significant spatial correlations when they are
accreted onto their hosts, those correlations should quickly decay due
to tidal gravity.  The exception to this argument is sub-substructure,
i.e. sub-subhalos that are gravitationally bound within larger subhalos.  In
general, however, sub-substructure comprises a very small fraction of
the total mass in substructure, and we therefore neglect it in our
calculations \citep{Diemand:07}.

We can therefore write down the substructure power spectrum as an integral over the subhalo mass function, weighted by the (Fourier transform of the) subhalo density profile, i.e.\ the 1-(sub)halo term:
\begin{equation}
P_\kappa(k) = \int \frac{dn}{dM} |\kappa_M(k)|^2 dM,
\label{P_halo}
\end{equation}
where $\kappa_M(k)$ is the Fourier transform of the convergence $\kappa_M$ provided by a subhalo of mass $M$, 
\begin{equation}
\kappa_M(k) = \int \kappa_M(r) e^{i k\cdot r} d^2 r = 
2\pi \int\kappa_M(r)J_0(kr) r\,dr,
\label{kappa_M}
\end{equation}
where the second equality holds for circularly symmetric
$\kappa_M(r)$.  Here, we make the flat-sky approximation, which is
quite accurate given the $\sim\,$arcsecond field of view
relevant for strong lensing. 

\begin{figure}
\begin{center}
\centering
\includegraphics[trim= 32 0 0 0, width=0.48\textwidth]{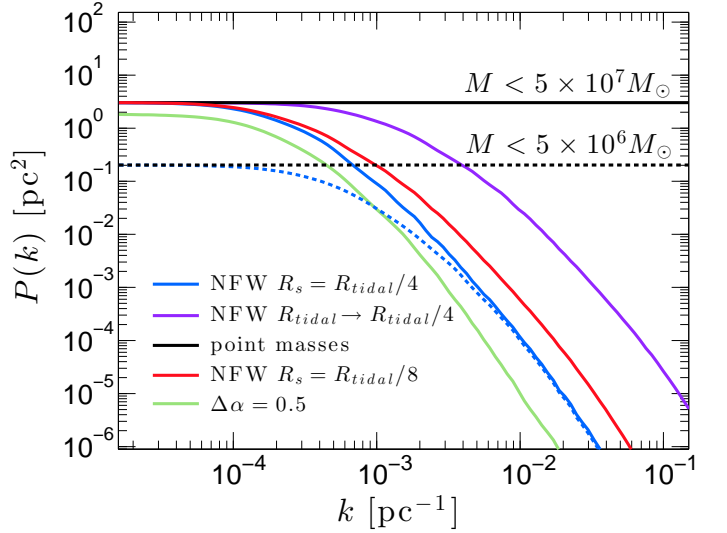}
\centering
\end{center}
\caption{ Power spectrum of projected density fluctuations from
  subhalos in the Via Lactea II (VL2) simulation.  Subhalo masses, sizes,
  and locations in the VL2 catalog are used to generate theoretical
  power spectra using Eqn.~(\ref{P_halo}).  The blue solid curve shows our fiducial model which includes subhalos with $M<5\times10^7 M_{\odot}$ with NFW profiles with $R_s=R_{tidal}/4$. The purple and red curves show the power spectrum when we alter the tidal radius, or the density profile ($R_s$) respectively. The solid black curve shows the power spectrum if the subhalos consist of point masses. The light-green curve shows the power spectrum when the slope of the mass function is altered by 0.5. The dotted lines show the power spectrum of subhalos with $M<5\times10^6 M_{\odot}$, for our fiducial model (blue), and for the point mass model. 
\label{f:f1}}
\end{figure}

Equation~(\ref{P_halo}) is instructive in understanding exactly what
aspects of the subhalo distribution control the form of the power
spectrum shown in Figure \ref{f:f1}.  For example, note that on large
scales (small wavenumber $k$), the substructure power spectrum
plateaus to a constant value.  The length scale above which $P(k)$
becomes flat corresponds to the sizes of the largest subhalos (compare
blue vs.\ purple curves in the Figure).  The amplitude of the power on
these large scales is determined by the total abundance of subhalos of
all masses, with a larger contribution from the most massive subhalos.
This can be understood
by inspecting Eqn.~(\ref{P_halo}).  Since $\kappa_M\propto M$, and
assuming a power-law mass function $dn/dM\propto M^{-\alpha}$, then
the integrand in Eqn.~(\ref{P_halo}) behaves as $M^{3-\alpha}$, which
is dominated by high masses for typical $\alpha\approx 2$.  Towards
smaller length scales,
the power spectrum changes shape, declining towards higher $k$.  The
shape of the power spectrum on these scales is affected by two
different terms: the internal profiles of massive halos, and the slope
of the subhalo mass function (through the connection of tidal radius to subhalo mass). 
Fig.~\ref{f:f1} illustrates the effects
of varying either of these properties.  Given a finite observable dynamic range, it may be
difficult to disentangle these two effects.  

\section{The Likelihood of the density power spectrum}
\label{sec:framework}
In this section, we describe the formalism for measuring the
substructure power spectrum from lensing measurements.  
Suppose we have observations $\bm O$ (e.g., surface brightness maps) and random measurement noise
$\bm N$ measured at $n$ pixels. At each pixel, there is also a random
deflection angle $\va$ coming from substructure.  We try to
model the observations with a model that has parameters $\vp$
describing the structure of the smooth lens potential and the background source emission.
Suppose that both the noise $\bm N$ and deflections $\va$ are
Gaussian random fields with probability: 
\begin{equation}
P(\bm N) =\frac{\exp\left(-\frac{1}{2}\bm N\cdot\CN^{-1}\cdot\bm N\right)}{(2\pi)^{n/2}|\CN|^{1/2}}
\end{equation}
where $\CN = \langle \bm N\, \bm N\rangle$ is the $n\times n$ noise covariance matrix, and similarly, 
\begin{equation}
\label{eq:P_alpha}
P(\va) =\frac{\exp\left(-\frac{1}{2}\va\cdot\Ca^{-1}\cdot\va\right)}{(2\pi)^{n}|\Ca|^{1/2}}
\end{equation}
where  $\Ca = \langle\va\,\va\rangle$ is the $2n\times2n$ covariance matrix for deflection angles.  Explicitly, 
\begin{eqnarray}
\label{eq:Pk}
\langle\alpha_i(\bm x)\,\alpha_j(\bm x+\bm r)\rangle &=& A_1(r)\delta_{ij} + A_2(r)\frac{r_i r_j}{r^2} \label{eq:Calpha} \\
A_1(r) &=& 4 \int |\kappa(k)|^2\frac{J_1(kr)}{kr} \frac{dk}{k} \nonumber \\  
A_2(r) &=& -4 \int |\kappa(k)|^2 J_2(kr) \frac{dk}{k}.  \nonumber
\end{eqnarray}
where we have used $\bm\nabla\cdot\va=2\kappa$.  
To estimate the likelihood for a given covariance given a set of measurements, we'll use Bayes' Theorem, which says that the likelihood for \Ca, $\CN$ is proportional to the likelihood for generating our observed measurements $\bm O_{\rm obs}$ given $\Ca$ and \CN:
\begin{eqnarray}
&&\mathcal{L}(\bm O_{\rm obs},\vp) = \int d^nN d^{2n}\!\alpha\, P(\bm N) P(\va)\,  \nonumber \\ 
&&\delta\!
 \left[\bm O_m(\vp) + \dOda\Delta\va + \bm N - \bm O_{\rm obs}\right]
P_{\rm p}(\vp) 
\end{eqnarray}
Here, $\vO_m(\vp)$ is the model prediction for parameter set $\vp$. 
Recall that $\vp$ includes parameters for both the smooth lens
  and the source emission.  In this work, we describe the source
  emission non-parametrically, as a pixelated map.  Because the source
  map has many degrees of freedom that are not fully constrained by
  the observations, regularization is required to avoid over fitting
  \citep[see e.g.,][]{Warren:03, Suyu:06}.  This regularization acts
  as a prior, $P_{\rm p}(\vp)$, which multiplies the above
  likelihood.  We use a Gaussian prior described by a covariance
  matrix \Cp,
\begin{equation}
P_{\rm p}(\vp) =\frac{\exp\left(-\frac{1}{2}(\vp-\vp_{\rm
      prior})\cdot\Cp^{-1}\cdot(\vp-\vp_{\rm prior})\right)}{(2\pi)^{n_p/2}|\Cp|^{1/2}},
\end{equation}
where $n_p$ is the number of parameters, and $\vp_{\rm prior}$ are fiducial parameters preferred by the prior.  Without loss of generality, we will set
$\vp_{\rm prior}=0$ to avoid confusion in the expressions below.

Assuming that the noise and substructure deflections are small, then the best-fitting parameters $\vp$ are always close to some fiducial parameter set $\vp_0$.  Taylor expanding, we have
\begin{equation}
\bm O_m({\vp}_0 + \Delta\vp) \approx \vO_0 + \dOdp\Delta \vp + \ldots,
\end{equation}

By marginalizing over the uncertain model parameters (smooth lens and source parameters) we can calculate the marginalized likelihood of the substructure covariance matrix. After a few lines of algebra we arrive at:
\begin{eqnarray}
\mathcal{L}(C_{\alpha}) = &&(|C_N| \, |C_{\alpha}| \, |C_p| |M|)^{-1/2}  e^{\frac{1}{2}B^\T \,M \,B}\nonumber \\
&& e^{-\frac{1}{2}(\Delta\vO^\T C_N^{-1} \Delta\vO+\vp_0 C_p^{-1} \vp_0)}
\label{eq:L}
\end{eqnarray} 
where 
\begin{equation}
M = \left[
\begin{array}{cc}
C_{\alpha}^{-1}+\dOda^\T C_N^{-1}\dOda& \dOda^\T C_N^{-1} \dOdp \\
\dOdp^\T C_N^{-1}\dOda & C_p^{-1} + \dOdp^\T C_N^{-1} \dOdp
\end{array}
\right] 
\end{equation}
and
\begin{equation}
B = \left[
\begin{array}{cc}
\Delta\vO^\T C_N^{-1}\dOda   &  \Delta\vO^\T C_N^{-1} \dOdp -  \vp_0^\T C_P^{-1}\\
\end{array}
\right] 
\end{equation}
Here $C_p$ is the prior covariance matrix (regularization matrix).
By Bayes' theorem, $\mathcal{L}(\Ca,\CN)\propto \mathcal{L}(\bm O_{\rm obs})$.  Thus,
by mapping out $\mathcal{L}(\bm O_{\rm obs})$ as a function of \Ca\ and \CN, we can determine the likelihood of the noise level and substructure power spectrum for each set of observations.

\section{Simulations}
\label{sec:simulations}

We generate mock observations of galaxies lensed by a halo (macro lens) and a population of subhalos, and use Equation (\ref{eq:L}) to map the likelihood of the amplitude of the power spectrum of the subhalo field, $P(k)$. 
The macro lens is modeled as a power-law elliptical mass
  distribution \citep{Barkana:98} plus two additional angular multipoles ($\cos 3\theta$
  and $\cos 4\theta$), along with external shear. The substructure population is modeled as a Gaussian random density field with a given power spectrum. 
To calculate the deflections due to the subhalo field, a map of substructure surface density is generated and the deflection angles are calculated in Fourier space using:
\begin{equation}
\tilde{\alpha} = \left( \frac{2 \, i \, k_x}{k^2}\tilde{\kappa} , \frac{2 \, i \, k_y}{k^2}\tilde{\kappa} \right)
\end{equation}
where $\tilde{\kappa}$ is the Fourier transform of the density field, and $k_x$ and $k_y$ are the Fourier coordinates. 
The surface density, $\kappa$, is generated by an inverse Fourier transformation of a map whose real and imaginary components are drawn at random to give rise to a desired power spectrum. To avoid periodicity and edge effects, we construct a subhalo density map that is approximately ten times larger than the $4\arcsec \times 4 \arcsec$ field of interest.  
The deflection angles due to the main lens and the substructure field
are then added together to predict the lensed images of a background
source.  The source consists of multiple (1-5) star-forming
clumps with a Gaussian light profile with FWHM of $\sim 300$ pc, distributed
over an area of $\sim1-2 \,$ kpc. 
As discussed in the previous section, we have used a pixelated
  grid to parameterize the background source emission, whose
  reconstruction is regularized using a Gaussian prior. 
  The source pixels covered an area of $3\times3$ kpc in the source plane with $40\times40$ pixels.  
Finally, Gaussian noise was then added to the lensed images, at levels based on the signal to noise of previous ALMA observations of lensed dusty sources.

For high-excitation molecular transitions it is expected that the
  source emission will be composed of a number of discrete clumps
  embedded in a larger structure, such as an exponential disk.  We
  used this structure to construct the source prior by calculating the
  power spectrum and covariance of such a clustered source model. 
The power spectrum of the source emission can be calculated in the same
way as our halo model approach for computing the lens substructure power spectrum.
 Suppose that we have $N_c$ clumps in our source galaxy, whose
 distribution within the galaxy has profile $U_c(r)$.  We normalize
 $U_c$ to have unit integral, $\int U_c (r) d^2r =1$.  Its Fourier
 transform is $U_c(k)$.  Clump $i$ has luminosity $L_i$ and profile
 $u_i(r)$, normalized to have unit integral, $\int u_i(r)d^2r =
 1$. Then the power spectrum of the source emission is proportional to 
\begin{equation}
P_{\rm src}(k) \propto \left[ \sum_i^{N_c}  L^2_i |u_i(k)|^2 + \sum_{i\neq j}^{Nc} L_iL_j |U_c(k)|^2 u^{*}_i(k) u_j (k) \right]
\label{sourcePk}
 \end{equation}  
To find the overall normalization of the covariance matrix we used the
method presented in \citet{Suyu:06} to maximize the evidence.  Note
that this clump model is only used to construct the covariance matrix
that regularizes the source reconstruction.  We do not assume that
the source is clumpy, but instead allow an arbitrary source emission.
This clump model is only used to regularize the pixelated source
reconstruction.  In agreement with previous work, we have found that
for data with sufficiently high signal-to-noise, the reconstruction
does not depend sensitively on the precise form of the regularization.

To map the likelihood of the power spectrum amplitude, we construct the deflection covariance matrix, \Ca, using equation (\ref{eq:Pk}). We use finite differencing to construct $\partial O/\partial p$ and  $\partial O/\partial \alpha$, and use the macro model (without including the substructure field) as the reference model, $O_0$.

 The size of these matrices and the computational cost of inversions grow very rapidly with increasing image resolution. For an image with $n\times n$ pixels, the deflection covariance matrix has a size of $2n^2 \times 2n^2$. Inversion of this matrix has a typical time complexity of  $\sim\mathcal{O}((2n^2)^{2.5})$.
In other words, doubling image resolution results in a $\sim32$-fold
increase in computational costs.  A single evaluation of the likelihood (which includes multiple inversions and determinant calculations) for a $50\times50$ image on a single CPU could take up to a few minutes. 
At higher resolutions, not only the likelihood evaluation becomes remarkably slower, retaining the data on single machine memory becomes unfeasible, with the overall size of matrices exceeding tens of GB for a $100 \times 100$ image. 
To overcome these obstacles we use the \textit{Elemental} package, an open-source \verb!C++! library for distributed-memory dense linear algebra \citep{Poulson:2013, Petschow:12}.

 In the simulations presented in this work we have assumed CCD data
 with uncorrelated noise in the images. This method, however, is
 equally applicable to interferometric data using the same fitting
 procedure as in \citet{hezaveh:13a} and \citet{hezaveh:13b}. In that
 case, the observables are the measured visibilities and, in the
 $uv$-space, the noise covariance matrix \CN\ is diagonal.

\section{Analysis and Discussions}
\label{sec:discuss}

\subsection{Substructure vs.\ other sources of fluctuations}

\begin{figure}
\begin{center}
\centering
\includegraphics[trim = 40 10 20 0 cm, width=0.435\textwidth]{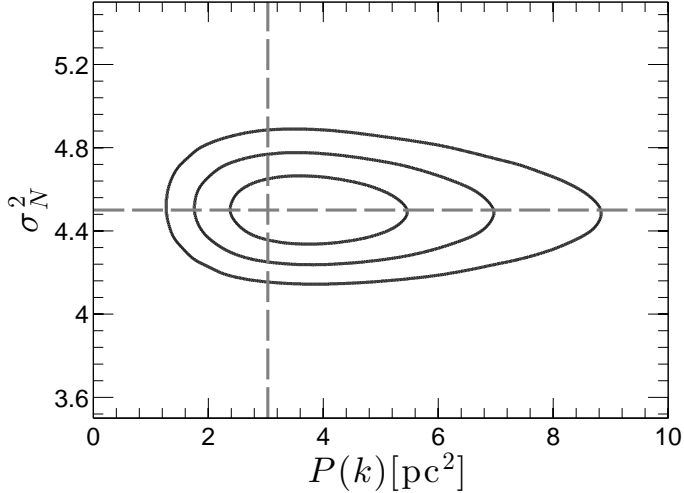}
\centering
\end{center}
\caption{ Joint-likelihood of noise and the amplitude of the power spectrum, mapped by evaluating Equation (\ref{eq:L})
using mock observations described in \S\ref{sec:simulations}, lensed
by a density field which includes substructure with a flat power spectrum.
The dashed lines show the true values which were used in the mock observation.
The input amplitude of the power spectrum is successfully recovered, with
little if any degeneracy between instrumental noise
and substructure fluctuations.
\label{f:kappa_noise}}
\end{figure}

Density perturbations produce random fluctuations in the lensed
observables.  Instrumental noise also produces random errors in the
observables, and naively we might expect that the effects of
measurement noise could be difficult to disentangle from the effects
of mass substructure.  Similarly, we might worry that fluctuations in
the source emission might also be degenerate with substructure
fluctuations in the lens mass distribution.
Figure \ref{f:kappa_noise} shows that this
is not the case.  The figure plots the joint likelihood (from
Equation \ref{eq:L}) as measured from simulated observations of sources
lensed by a main halo and a density field with a flat, white noise
power spectrum.  As described above, we have marginalized over a
pixelated source emission grid.
As is apparent, there is little if any degeneracy
between instrumental noise and the amplitude of the density power
spectrum.  The density field in this simulation only contains modes
between $\sim0.04-0.4\ {\rm kpc}^{-1}$, which roughly cover the range
of modes where the power spectrum is expected to be flat (see Figure \ref{f:f1}).
Repeating this procedure for various mock observations indicates that
the two parameters are non-degenerate over the entire range that we
have simulated.  We again stress that this calculation marginalizes
over the source emission, so the fact that we recover the input power
spectrum implies that substructure in the lens mass is not degenerate
with clumpiness in the source emission.
We have evaluated this likelihood using different source priors with different parameters (i.e. $U_c$ and $u$ in Equation \ref{sourcePk}) and found that for high signal to noise observations the precise form of the prior does not appear to significantly affect the reconstruction.

Given the lack of any degeneracy between measurement noise and lens
substructure, and the fact that the noise
properties of most observations can be precisely quantified, in the
rest of this work we do not map the likelihood along the noise
dimension, assuming that the noise rms is accurately known.

\subsection{Sample variance}

\begin{figure}
\begin{center}
\centering
\includegraphics[trim = 38 10 20 0 cm, width=0.43\textwidth]{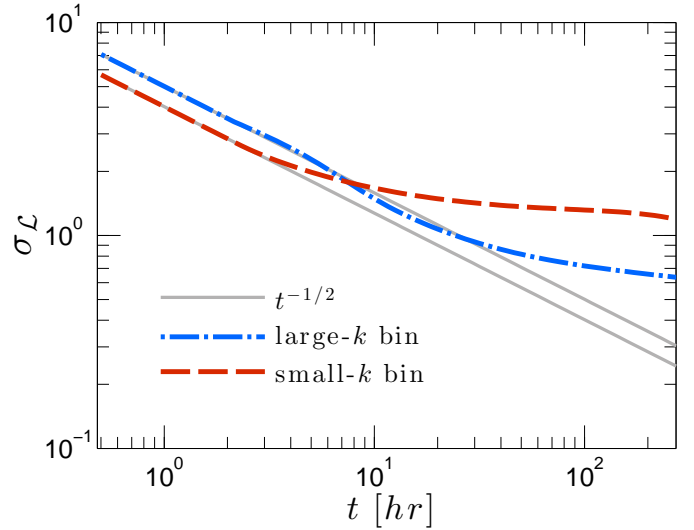}
\centering
\end{center}
\caption{ The rms of the likelihood of the amplitude of the power
  spectrum as a function of observing time. For purely noise dominated
  observations, these curves should scale as $t^{-1/2}$, as seen for
  observations $\lesssim10$ hours.  For longer observations, the errors
  are dominated by sample variance. The contribution of sample
  variance is larger for modes with smaller wavenumber $k$, (red)
  compared to the more numerous modes at higher frequency (blue).
\label{f:samplevariance}}
\end{figure}

 Next, we investigate how the errors of the power spectrum measurement
scale with signal to noise.  Our results suggest that the power
spectrum uncertainty becomes sample variance dominated for observations with very high signal to noise. The red and blue dashed curves in Figure \ref{f:samplevariance} show the rms of likelihood as a function of observing time for continuous observations. The blue dashed curve corresponds to the rms for modes with larger wavenumber than the ones for the red curve. As seen in the figure, at higher wavenumber (blue curve), each bin contains a larger sample of modes, helping to reduce sample variance. 
The gray lines show the $t^{-1/2}$ scaling expected for measurements that are purely noise dominated. To overcome the limitation imposed by sample variance when higher precision measurements are needed, we can combine measurements from many lenses, assuming that they are different realizations of the same process. This approach, however, requires a careful analysis of the selection methods and the scaling of the subhalo properties with that of the host halo \citep{xu:13}.

\begin{figure*}
\begin{center}
\centering
\includegraphics[trim= -1 15 0 15, width=0.6\textwidth]{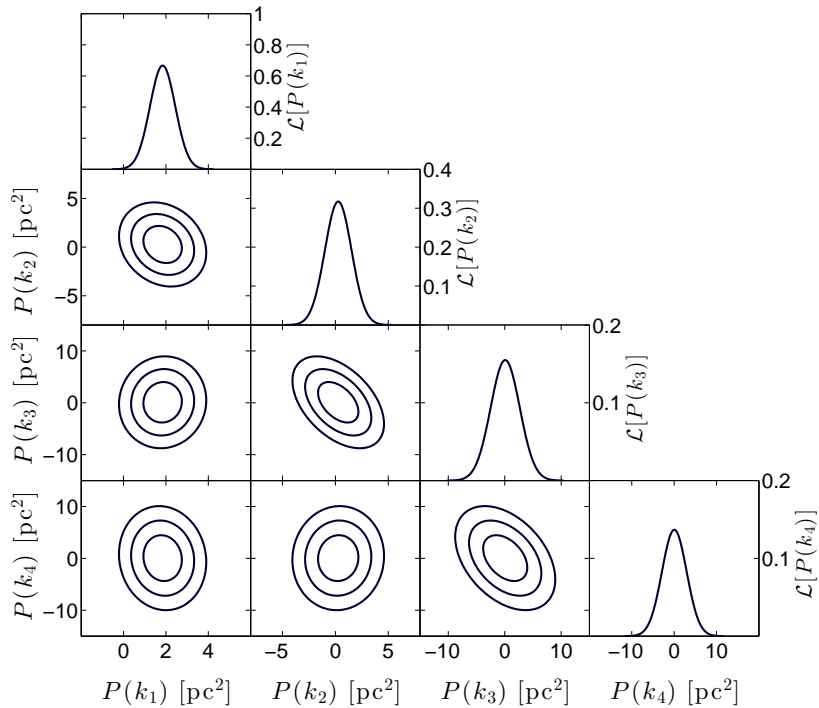}
\centering
\end{center}
\caption{ Fisher matrix forecasts for $P(k)$ errors for a 10 hour long
  observation.  Here we have parametrized the power spectrum with
  bandpowers in 4 bins, $k_1$ to $k_4$, corresponding to modes in the ranges
  $0.04-0.4$,  $0.4-0.8$,  $0.8-1.2$, and   $1.2-1.7$  kpc$^{-1}$
  respectively.  Negative allowed values of $P(k)$ are, of course,  an
  artifact of the Fisher matrix approximation to the full likelihood.
\label{f:fisher}}
\end{figure*}

\subsection{Measuring the shape of $P(k)$}

So far, we have discussed measurements of flat (white noise) power
spectra.  This power spectrum arises from populations of point masses (e.g. primordial black holes).
As discussed in section \ref{sec:power}, realistic dark matter subhalos,
with finite sizes and smooth density profiles, give rise to power
spectra that are flat on large scales (small $k$) and fall off at
large wavenumbers (Figure \ref{f:f1}).  To measure a power spectrum
with an arbitrary shape, we can parameterize $P(k)$ in Equation
\ref{eq:Pk} as a function of $k$ when constructing \Ca, and fit for
the free parameters using Equation \ref{eq:L}.  For example, we could
assume that the power spectrum is piecewise flat in discrete $k$ bins,
and then use the bandpowers in those bins as the free parameters.  For
sufficiently narrow bins, we can approximate arbitrary power spectra
this way. Noting that the power is relatively flat over $0.04-0.4\,{\rm kpc}^{-1}$, we choose a linear binning in which the first bin covers the range of $k$ modes over the flat part of the power spectrum, and the other bins measure the fall of the power spectrum at high $k$. With this choice of binning, a measurement of the amplitude of the modes in the first bin gives power over the flat part of the power spectrum, revealing the total abundance of all low-mass subhalos.

For $n$ bins over the available range of $k$'s, the power spectrum is defined in an $n$-dimensional parameter space. 
Since the likelihood evaluations are computationally expensive, we use a Fisher analysis to forecast the size of the errorbars and the degeneracies between power at different scales for different observing conditions. The input power spectrum of the subhalo density field in the mock observations is set to be consistent with the Via Lactea II (VL2) simulation: the positions, masses, and tidal radii of the subhalo are taken from the publicly available VL2 catalogue\footnote{http://www.ucolick.org/$\sim$diemand/vl/data.html} \citep{ViaLactea2} and the subhalos are given a truncated NFW profile with $R_s=R_{tidal}/4$. Figure \ref{f:fisher} shows an example of the parameter covariance (amplitudes in four bins) for a simulated observation.

\begin{figure}
\begin{center}
\centering
\includegraphics[trim= 0 30 0 -10, clip, width=0.51\textwidth]{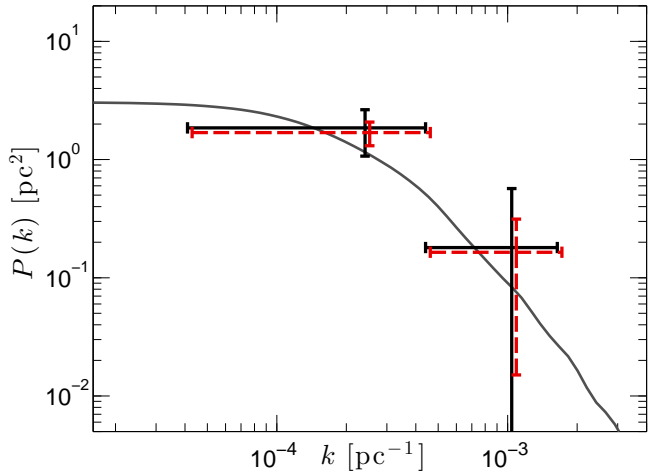}
\centering
\end{center}
\caption{Forecast for a measurement of the power spectrum of subhalos
  with $M<5\times 10^7 M_{\odot}$ for a 10-hour long observation (black errorbars) of a single source with
  ALMA, assuming an observed source continuum flux of 
  50 mJy at 850 $\micron$.
   The first bin, with a significance of $\sim3\sigma$, indicates the abundance of all subhalos in the
  main dark matter halo. Deeper observations ($\sim 40$ hr),
  combination of all the modes at higher $k$,  and more favorable
  conditions (smaller source size) could allow a measurement of the
  break in the power spectrum at higher $k$ (red errorbars). The underlying power spectrum is the fiducial model of Figure 1 (blue curve in Figure 1).
\label{f:errorpredictions}}
\end{figure}

Figure \ref{f:errorpredictions} shows the errorbars of two bins for a
signal to noise comparable to a 10-hr long ALMA observation of bright
lensed dusty galaxies. This results in a detection of the power in the
first bin ($\sim3 \sigma$) revealing the total abundance of
subhalos. On smaller scales, the predicted power spectrum
falls too rapidly and this observation can only put an upper bound on
the high-$k$ amplitude.  This upper limit, however, may be adequate to
indicate a break in the power spectrum.  An observation approximately
4 times longer (or involving $4$ different lens systems) could measure
the power over this regime.

\subsection{Non-Gaussianity}

\begin{figure}
\begin{center}
\centering
\includegraphics[trim= 20 10 0 0, width=0.48\textwidth]{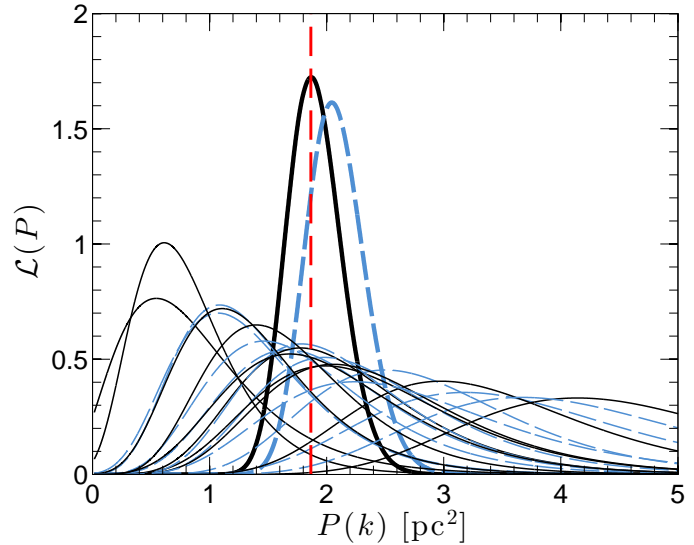}
\centering
\end{center}
\caption{
The thin curves show the recovered likelihoods of the amplitude of the power spectrum in ten realizations of a Gaussian (solid black) and non-Gaussian (dashed blue) density field.
In the case of blue curves, the subhalo density field is
 non-Gaussian, but incorrectly treated as a Gaussian field. 
The thick black solid curve show the combined likelihood that results from multiplying the ten Gaussian likelihoods together. The thick blue dashed curve shows the same for the non-Gaussian density maps.
The deviation of the blue dashed curve from the solid curve ($\sim7\%$)
 results from biasing the likelihood estimator due to non-Gaussianity
 in the substructure field.
 Repeating the test for 100 realizations yields consistent results,
 with a corresponding improvement in the joint likelihood.
\label{f:nongaussianity}}
\end{figure}

So far, we have assumed that the subhalo density field could be treated as a Gaussian random field. 
In reality, the substructure field is not Gaussian distributed. The
non-Gaussianity mainly arises from the few most massive subhalos. To reduce this non-Gaussianity, it is important to be able to detect and remove the effect of the most massive subhalos with low number densities. The power spectra used for simulations in this work were calculated for subhalos with $M<5\times 10^7 M_{\odot}$, assuming that subhalos with masses larger than this limit could be detected individually using a direct lens modeling approach  \citep[e.g.][]{vegetti:12, hezaveh:13a}. 
To estimate how much the remaining non-Gaussianity in the density
field biases our results, we performed 100 simulations of Gaussian and
non-Gaussian substructure density fields. The non-Gaussian maps were
generated with subhalo masses and numbers taken from the Via Lactea II
catalogue. After mapping the likelihood of the power spectrum
amplitude for each simulation, we multiplied the hundred likelihoods
together for the Gaussian and non-Gaussian case. Figure
\ref{f:nongaussianity} shows a subset of the resulting
likelihoods. The modes depicted in the plot correspond to the first
bin of Figure \ref{f:errorpredictions}, covering the flat part of the spectrum.
This figure shows that in the case of non-Gaussian density fields, the
true value is about 7\% biased. Although this is a biased measurement,
given that we currently do not know the value of this power spectrum
to any precision, a 7\% biased measurement is valuable. However, it is
also possible to avoid this bias, in principle.

If we can assume a known profile for subhalos (e.g.\ truncated NFW), then we can generate Monte
Carlo realizations of the non-Gaussian density field and constrain
substructure properties using the method of
\cite{dalal:02}.  
Additionally one can perform such analysis for different density profiles and marginalize over their parameters.
That method, however, is considerably more
computationally intensive than the Gaussian likelihood estimator
discussed in this paper.

\section{Conclusions}
\label{sec:conclusions}

In this paper, we presented a framework for measuring the power
spectrum of the substructure density field using observations of
strong gravitational lenses.  We showed how the amplitude and shape of
the power spectrum is related to the abundance, density profile, and
mass function of subhalos.  Using mock observations, we tested the
method, successfully recovering the input parameters and showed that
if dark matter halos host large populations of subhalos consistent
with CDM simulations, this power spectrum could be measured with
near-future observations.

We found that $\sim10$ hour long ALMA observations of a single lensed submm 
source should be capable of detecting the amplitude of the substructure power spectrum 
at $\gtrsim3\sigma$ significance.
In our calculations, we have marginalized over any uncertainty in the
`macro' model describing the smooth mass distribution of the main
lens.  These macro parameters are completely degenerate with the
longest wavelength modes of the substructure field, implying that the
lowest $k$ modes will be unconstrained.  Fortunately,
however, such unconstrained modes are few in number, and their
degeneracy does not significantly impact on our ability to measure low
$k$ bandpowers, as long as sufficiently wide $k$-bins are employed, as
illustrated in Figs.~\ref{f:fisher} and \ref{f:errorpredictions}.

The power spectrum measured from lensing observations may be directly
compared to results of numerical simulations.  Although we have
interpreted the power spectrum in terms of the abundance of dark
matter subhalos, the same quantity measured by lensing may also
directly be measured in simulations without resorting to catalogs of
subhalos.  This may prove to be a useful approach, since subhalo
properties are notoriously difficult to measure in simulations:
different subhalo finders applied to the same simulations can
sometimes produce discrepant results, depending on the subhalo
definitions and parameter choices adopted by the various finders \citep{Onions:12}.
By directly measuring substructure power spectra from simulations,
uncertainties in the definition of subhalos may be circumvented.

Lastly, we note that our results depend on our choice of parameters in
our simulations.  Wherever possible, we have attempted to be
conservative in our choices.  We have generated macro lenses and
source brightnesses consistent with existing low-resolution imaging of
sub-mm lenses from ALMA \citep{hezaveh:13b, vieira:13}.  The main uncertainty in
our simulations is the unknown number and size distribution of
star-forming clumps in the source galaxies.  To be conservative, we
have assumed clump sizes of $\sim 200$ pc, the upper limit placed by current observations \citep[e.g.][]{swinbank:10}. If the source clumps are smaller in
reality than we have assumed, then our forecasted constraints on the
high-$k$ power spectrum can improve significantly.

\acknowledgements{ This work was supported by NASA under grant
  NNX12AD02G.  
YH thanks the \emph{Elemental} team, Jack Poulson and Jeff Hammond for tremendous support. YH acknowledges very useful and constructive discussions with Ryan Keisler, Phil Marshall, Roger Blandford, James Bullock, and David Spergel. 
This work was initiated at the Aspen Center for Physics, which is supported by NSF grant 1066293. We thank Calcul Quebec for providing the computing resources that we used for a part of this research.}

\bibliographystyle{apj}

\end{document}